\documentstyle[12pt,times,fullpage,epsf]{article}
\begin{document}
\begin{center}
{\bf Correlated Polarons in Dissimilar Perovskite Manganites}

\vspace{0.25in}

C.S. Nelson$^{1}$, M. v. Zimmermann$^{1}$, J.P. Hill$^{1}$, Doon Gibbs$^{1}$, V. Kiryukhin$^{2}$, T.Y. Koo$^{2,3}$, S-W. Cheong$^{2,3}$, D. Casa$^{4}$, B. Keimer$^{5}$, Y. Tomioka$^{6}$, Y. Tokura$^{6,7}$, T. Gog$^{8}$, and C.T. Venkataraman$^{8}$

\vspace{0.25in}

$^{1}$Department of Physics, Brookhaven National Laboratory, Upton, NY 11973-5000

$^{2}$Department of Physics and Astronomy, Rutgers University, Piscataway, NJ  08854

$^{3}$Bell Laboratories, Lucent Technologies, Murray Hill, NJ  07974

$^{4}$Department of Physics, Princeton University, Princeton, NJ  08544

$^{5}$Max-Planck-Institut f\"{u}r Festk\"{o}rperforschung, 70569, Stuttgart, Germany

$^{6}$Joint Research Center for Atom Technology (JRCAT), Tsukuba 305-0033, Japan

$^{7}$Department of Applied Physics, University of Tokyo, Tokyo 113-0033, Japan

$^{8}$CMC-CAT, Advanced Photon Source, Argonne National Laboratory, Argonne, IL  60439

\vspace{0.125in}

\today

\end{center}

\vspace{0.125in}

We report x-ray scattering studies of broad peaks located at a (0.5 0 0)/(0 0.5 0)-type wavevector in the paramagnetic insulating phases of La$_{0.7}$Ca$_{0.3}$MnO$_{3}$ and Pr$_{0.7}$Ca$_{0.3}$MnO$_{3}$.  We interpret the scattering in terms of correlated polarons and measure isotropic correlation lengths of 1--2 lattice constants in both samples.  Based on the wavevector and correlation lengths, the correlated polarons are found to be consistent with CE-type bipolarons.  Differences in behavior between the samples arise as they are cooled through their respective transition temperatures and become ferromagnetic metallic (La$_{0.7}$Ca$_{0.3}$MnO$_{3}$) or charge and orbitally ordered insulating (Pr$_{0.7}$Ca$_{0.3}$MnO$_{3}$).  Since the primary difference between the two samples is the trivalent cation size, these results illustrate the robust nature of the correlated polarons to variations in the relative strength of the electron-phonon coupling, and the sensitivity of the low-temperature ground state to such variations.

\pagebreak

The wide variety of ground state phases exhibited by the perovskite manganites (R$_{1-x}$M$_{x}$MnO$_{3}$ where R and M are trivalent rare-earth and divalent alkaline cations, respectively) originates in the interplay of the charge, orbital, lattice, and spin degrees of freedom.  A related manifestation of this interplay is the colossal magnetoresistance (CMR) effect \cite{cmr}, a phenomenon that has caused a resurgence of interest in the perovskite manganites due to its potential for technological applications.  Recent work has focused on the role of the electron-phonon interaction, which has been used to supplement the double exchange interaction \cite{zener}, and is believed to be a necessary ingredient for modeling the temperature dependence of the magnetic and transport behavior of CMR materials \cite{roder,millis2}.  

In CMR materials, strong electron-phonon coupling results in the formation of localized charge carriers with associated lattice distortions--- or polarons--- in the paramagnetic insulating phase.  Early evidence of polarons was obtained from transport measurements \cite{hundley,jaime}, from which a high-temperature/small polaron and low-temperature/large polaron phenomenology was hypothesized.  Studies using local probes \cite{billinge,shengalaya} supported this view of the high-temperature, polaronic behavior, and more recently, interest has turned to consideration of polaron-polaron interactions in this temperature regime.  One theoretical model, proposed by Alexandrov and Bratkovsky \cite{alexandrov}, involves paired polarons, or bipolarons.  In this model, strong electron-phonon coupling at high temperatures binds polarons into immobile pairs in a singlet state, with two holes localized on a single oxygen ion.  Then, as the temperature is reduced toward T$_{c}$, the ferromagnetic exchange coupling increases in strength and the bipolarons are broken apart, leaving single polarons in the ferromagnetic metallic phase.  Evidence consistent with bipolarons--- based on transport measurements of polycrystalline La$_{5/8}$Ca$_{3/8}$MnO$_{3}$ \cite{kim} and films of La$_{0.75}$Ca$_{0.25}$MnO$_{3}$ and Nd$_{0.7}$Sr$_{0.3}$MnO$_{3}$ \cite{zhao}--- has recently been reported, but additional data using more direct techniques are clearly desirable.

X-ray and neutron scattering experiments can make an important contribution to studies of polarons since they are directly sensitive to both the polarons and their correlations.  Recently, these techniques were applied to the perovskite manganite (Nd$_{0.125}$Sr$_{0.875}$)$_{0.52}$Sr$_{0.48}$MnO$_{3}$ \cite{shimo} and the layered manganite La$_{1.2}$Sr$_{1.8}$Mn$_{2}$O$_{7}$ \cite{doloc}.  In both materials, diffuse scattering around the Bragg peaks and well-resolved peaks at incommensurate wavevectors were observed in the high-temperature phases, and both types of scattering were found to disappear as the samples were cooled through T$_{c}$.  The two scattering components were attributed to the presence of polarons--- the diffuse scattering from single polarons and the resolved peaks from polaron correlations--- implying that both polarons and polaron correlations were absent in the ferromagnetic metallic phases.  Very recently, neutron scattering studies of the La$_{1-x}$Ca$_{x}$MnO$_{3}$ system \cite{dai,adams} have reported similar results.  In this case, however, the correlations were observed at a commensurate wavevector of (0.25 0.25 0), in cubic notation.

As discussed above, the important parameter controlling the polaron behavior in CMR materials is the electron-phonon coupling.  What is missing in the work reported to date is a systematic study of the polarons as a function of the relative strength of this coupling.  In this paper, we take a step in this direction by reporting x-ray scattering studies of two identically-doped perovskite manganites, which have different trivalent cations.  To first order, such substitution does not affect the charge, orbital, or spin degrees of freedom but only alters the lattice degree of freedom, due to the difference in trivalent cation size.  One result of this is a change in the relative strength of the electron-phonon coupling.  The systems chosen for this study are La$_{1-x}$Ca$_{x}$MnO$_{3}$ and Pr$_{1-x}$Ca$_{x}$MnO$_{3}$, with a doping of x = 0.3.

La$_{0.7}$Ca$_{0.3}$MnO$_{3}$ and Pr$_{0.7}$Ca$_{0.3}$MnO$_{3}$ are both paramagnetic insulators at room temperature.  Upon cooling, these two materials go through transitions into completely different low-temperature phases--- a ferromagnetic metallic phase \cite{schiffer} and an antiferromagnetic, charge and orbitally ordered, insulating phase \cite{jirak,tomioka}, respectively.  The contrasting transport behavior can be seen in the resistivity measurements displayed in Figure 1.  The origin of the difference in the low-temperature phases is believed to be the decrease in cation radius from La to Pr, which is $\sim$3\%.  The smaller Pr ions produce a larger distortion of the Mn--O--Mn bond angles away from 180$^{\circ}$ (156.4$^{\circ}$ for Pr versus $\sim$160.7$^{\circ}$ for La \cite{hangs}), resulting in a smaller bandwidth and elastic modulus.  Both of these effects increase the relative strength of the electron-phonon coupling \cite{millis1,roder,egami}.

Here, we report x-ray scattering studies of the polarons in the paramagnetic insulating phases of these dissimilar perovskite manganites.  We find that the high-temperature behaviors are insensitive to the variation in the relative strength of the electron-phonon coupling between the two samples.  Specifically, in both samples, we observe broad, commensurate peaks with similar correlation lengths of 1--2 lattice constants, and an absence of diffuse scattering that could be attributed to single polarons.  We interpret the peaks in terms of correlated polarons, and based on the wavevector and correlation length, we find them to be consistent with a bipolaron model with the structure of a CE-type orbital order domain.  The difference in the relative strength of the electron-phonon coupling in the two samples is only manifested upon cooling through the respective transition temperatures.  In La$_{0.7}$Ca$_{0.3}$MnO$_{3}$, the broad peak decreases in intensity as the sample becomes ferromagnetic metallic, while in Pr$_{0.7}$Ca$_{0.3}$MnO$_{3}$, the peak narrows and increases in intensity as the sample enters a CE-type orbital order domain state.

The La$_{0.7}$Ca$_{0.3}$MnO$_{3}$ and Pr$_{0.7}$Ca$_{0.3}$MnO$_{3}$ single-crystals used in this study were grown by floating zone techniques at Bell Laboratories and JRCAT, respectively.  The x-ray scattering measurements were carried out on beamline X22C at the National Synchrotron Light Source (NSLS) and beamline 9ID at the Advanced Photon Source.  The incident photon energy was set near the Mn K edge (6.539 keV).  Both samples were determined to be fully twinned, with (110)/(002)-oriented surface normals (in orthorhombic, {\it Pbnm}, notation) and mosaic widths of $\sim$0.2$^{\circ}$ (FWHM).  For simplicity, all reflections discussed here are referenced using the (110) surface normal direction.  

We begin with the La$_{0.7}$Ca$_{0.3}$MnO$_{3}$ sample.  At temperatures above the metal-insulator transition temperature ($\sim$252 K), broad peaks with ordering wavevectors of (0.5 0 0) and (0 0.5 0) and peak intensities of $\sim$20 counts/s (on beamline X22C at the NSLS) were observed.  Note that twinning of the sample and the width of the diffuse peaks make it impossible to determine whether or not there is a unique ordering wavevector.  As the sample was cooled through the transition temperature into the ferromagnetic metallic phase, the peaks abruptly decreased in intensity (see inset to Figure 2(a)).  Two temperature snapshots of this behavior at 260 K and 220 K are shown in Figure 2.  

Reciprocal space mesh scans around the (220) Bragg peak were carried out at temperatures of 260 and 220 K.  No change in the intensity or shape of the x-ray diffuse scattering that would suggest the presence of single polarons was observed.  This lack of diffuse scattering appears to be at odds with the neutron scattering work reported by Dai {\it et al.} \cite{dai} and Adams {\it et al.} \cite{adams}, and could indicate that the diffuse scattering observed in these experiments is of magnetic origin, though it is also possible that evidence of the presence of single polarons is beyond our detection limits \cite{caveat}.

The temperature dependence of the diffuse peak at (1.5 2 0) was studied using reciprocal space scans along H and K.  Fitting the data to a Lorentzian-squared lineshape provides information about the correlation lengths along the orthorhombic {\it a} and {\it b} directions.  The correlation lengths are defined as $\textstyle\xi_{a}\equiv\frac{a}{2\pi\Delta H}$ and $\textstyle\xi_{b}\equiv\frac{b}{2\pi\Delta K}$, where $a$ and $b$ are the lattice constants and $\Delta H$ and $\Delta K$ are the half-width-at-half-maximum (HWHM) values of the diffuse peaks along H and K, respectively.  Above the transition temperature, the correlation lengths of the diffuse peak were observed to be independent of temperature and isotropic, with a magnitude of 1--2 lattice constants in both directions.

Turning now to Pr$_{0.7}$Ca$_{0.3}$MnO$_{3}$, the x = 0.3 doping in this system sits near the phase boundary between the CE-type charge and orbitally ordered, antiferromagnetic, insulating phase (T$_{co}\approx$ 220 K, T$_{N}\approx$ 150 K) and a ferromagnetic insulating phase (T$_{c}\approx$ 140 K) \cite{tomioka}.  Because of the close proximity to different low-temperature phases--- coupled with recent reports of phase separation at this doping \cite{martin,radaelli}--- we first carried out measurements at low temperatures ($\sim$100 K) with a high-resolution Ge(111) analyzer in order to characterize the ordered phase.  Three ordered phases associated with at least two different crystallographic phases were observed.  Two of these ordered phases are consistent with a CE-type charge and orbital order--- with (100)/(010) and (0.5 0 0)/(0 0.5 0) wavevectors, respectively--- while the third exhibited only a (100)/(010) ordering wavevector.  The distinct nature of the three phases was determined via measurements of their respective temperature dependences.  The two CE-type phases were found to have ordering transitions at 130 and 200 K, and the ordering in the third phase persisted up to room temperature.  For the purpose of this paper, we focus on the CE-type phase with the higher transition temperature in what follows, since its behavior is consistent with previous studies of Pr$_{0.7}$Ca$_{0.3}$MnO$_{3}$ \cite{tomioka}.  A complete analysis of the phase separation in Pr$_{0.7}$Ca$_{0.3}$MnO$_{3}$ will be presented in a future paper \cite{nelson}.

The (1.5 2 0) orbital order peak was measured while warming the sample up to the ordering temperature of $\sim$200 K.  At low temperatures, the peak width was measured using a Ge(111) analyzer, and a correlation length of 170 $\pm$ 20 $\rm \AA$ was determined.  As the sample temperature was increased through the ordering temperature, the (1.5 2 0) intensity decreased dramatically, as shown in Figure 3.  The peak was also observed to broaden rapidly, reaching a value corresponding to a correlation length of 1--2 lattice constants at high temperatures.  

As was done with La$_{0.7}$Ca$_{0.3}$MnO$_{3}$, reciprocal space mesh scans were also carried out near the (2 2 0) Bragg peak at temperatures both above and below the Pr$_{0.7}$Ca$_{0.3}$MnO$_{3}$ transition temperature.  Again, no evidence of diffuse scattering that could be attributed to the presence of single polarons was observed \cite{caveat}.

To quantitatively compare the two samples, the sets of data measured along H were analyzed by fitting to a double Lorentzian-squared lineshape \cite{nojust}.  This procedure enabled the diffuse \linebreak (1.5 2 0) peak to be separated out from the tails of the (2 2 0) Bragg peak.  The results of these fits are summarized in Figures 4 and 5, showing the peak intensity and the HWHM, respectively, versus reduced temperature.  In each case, the reduced temperature, $t$, is defined as $\textstyle\frac{T-T_{\rho/co}}{T_{\rho/co}}$, where $T_{\rho/co}$ is the metal-insulator (La$_{0.7}$Ca$_{0.3}$MnO$_{3}$) or charge and orbital order (Pr$_{0.7}$Ca$_{0.3}$MnO$_{3}$) transition temperature.

With regard to both intensities and correlation lengths, the scattering due to correlated polarons in the two samples exhibits strikingly similar behavior above the respective transition temperatures.  That is, the intensities appear to decrease gradually with increasing temperature, and the correlation length remains on the order of 1--2 lattice constants up to the highest temperature studied.  The energy dependence of the scattered intensity at the (1.5 2 0) peak was also found to be similar in the two samples.  Specifically, the intensity exhibited a sharp drop near the Mn K edge, consistent with normal charge scattering, and indicating that the scattering at this Q is primarily due to lattice distortions in both samples.

In the paramagnetic insulating phases of both samples, the magnitude of the wavevector, the short correlation length, and the absence of diffuse scattering around the Bragg peaks \cite{caveat} suggest an additional conclusion--- that the polarons are actually bipolarons, with a structure such as that of an orbital order domain in the CE-type phase.  The proposed bipolaron structure is displayed in Figure 6.  It consists of neighboring orthorhombic unit cells along the [110] direction, with Mn$^{4+}$ ions situated between orbitally-ordered Mn$^{3+}$ ions.  This CE-type bipolaron can also be viewed as a ferromagnetic zigzag, which is consistent with the observation of ferromagnetic fluctuations in Pr$_{1-x}$Ca$_{x}$MnO$_{3}$--- for the nearby doping range of x$\simeq$0.35--0.5 \cite{kajimoto}--- and the absence of antiferromagnetic correlations in La$_{0.7}$Ca$_{0.3}$MnO$_{3}$ \cite{adams}.  We also note that the presence of such ferromagnetic zigzags was recently proposed based on transport measurements of polycrystalline La$_{5/8}$Ca$_{3/8}$MnO$_{3}$ \cite{kim}.  The proposed CE-type bipolaron structure is therefore consistent with known experimental results, but it is not unique and alternative structures could be constructed.  We note that our results are inconsistent, though, with the theoretical model described earlier \cite{alexandrov}, in that two holes on the same ion would result in a much shorter correlation length than the measured value of 1--2 lattice constants.  In addition, a competing structure that has recently been proposed \cite{kilian,mizokawa}--- the orbital polaron, in which a Mn$^{4+}$ ion is surrounded by six Mn$^{3+}$ ions with their occupied e$_{g}$ (3d$_{z^{2}-r^{2}}$) orbitals pointing toward the central Mn$^{4+}$ ion--- is inconsistent with the wavevector reported here.  To summarize, we find the evidence in favor of CE-type bipolarons to be persuasive, but the discrepancy observed between neutron and x-ray scattering measurements of the diffuse scattering, or lack thereof, in the tails of the Bragg peaks remains to be reconciled.

The difference in the relative strength of the electron-phonon coupling for the two samples is manifested only at low temperatures, below the respective transition temperatures.  As the \linebreak La$_{0.7}$Ca$_{0.3}$MnO$_{3}$ sample is cooled through the metal-insulator transition, the (1.5 2 0) intensity decreases with no change in the correlation length.  This is consistent with the bipolaron picture described above, in which localization of correlated charge carriers is destroyed as the sample becomes conducting.  In contrast, the (1.5 2 0) intensity grows and the width decreases in Pr$_{0.7}$Ca$_{0.3}$MnO$_{3}$ as the sample goes through the charge and orbital order transition temperature.  A plausible scenario for the transition in this case is that the number of bipolarons increases until they begin to coalesce, eventually forming orbitally ordered domains.  This increase in the number of bipolarons could perhaps be driven by charge order fluctuations, consistent with the phenomenology reported for the x = 0.4 and 0.5 dopings in this system \cite{martin1,martin2}.  The observation of low-temperature, resolution-limited, charge order peaks using the high-resolution Ge(111) analyzer at the (030) peak in the Pr$_{0.7}$Ca$_{0.3}$MnO$_{3}$ sample lends support to this scenario. 

In conclusion, we have studied two CMR materials that differ only with respect to their trivalent cation species.  A comparison of the high- and low-temperature behaviors underscores the importance of the electron-phonon coupling in such materials.  In the high-temperature, paramagnetic, insulating phases, the behaviors of the polarons are found to be remarkably similar.  That is, as the temperature is reduced toward the transition temperature, a slight increase in the number of correlated polarons is observed at the same (0.5 0 0)/(0 0.5 0)-type ordering wavevector, and with the same isotropic correlation length of 1--2 lattice constants in the two materials.  The correlated polarons are consistent with a CE-type bipolaron structure, and the similar behaviors in the two samples suggest that these correlated polarons are robust with respect to variations in the relative strength of the electron-phonon coupling.

The similarity in the behavior of the two samples breaks down upon cooling through their respective transition temperatures, as the small difference in the size of the trivalent cations--- reflected in the relative strength of the electron-phonon coupling--- becomes important.  Below their transition temperatures, La$_{0.7}$Ca$_{0.3}$MnO$_{3}$ becomes a ferromagnetic metal and Pr$_{0.7}$Ca$_{0.3}$MnO$_{3}$ becomes a charge and orbitally ordered insulator.  In La$_{0.7}$Ca$_{0.3}$MnO$_{3}$, the collapse of the bipolarons can be attributed to the onset of ferromagnetic ordering, at which the energy gain associated with the ferromagnetic state overcomes the localization of the pairs of charge carriers caused by electron-phonon coupling.  In Pr$_{0.7}$Ca$_{0.3}$MnO$_{3}$, which has the stronger electron-phonon coupling, cooling through the transition temperature results instead in an ordering of the correlated polarons.  The driving mechanism for this ordering is as yet unknown, but charge ordering is a candidate and will be the subject of future experiments.  

The work at Brookhaven, both in the Physics Department and at the NSLS, was supported by the U.S. Department of Energy, Division of Materials Science, under Contract No. DE-AC02-98CH10886, and at Princeton University by the National Science Foundation under Grant No. DMR-9701191.  Work at the CMC beamlines is supported, in part, by the Office of Basic Energy Sciences of the U.S. Department of Energy and by the National Science Foundation, Division of Materials Research.  Use of the Advanced Photon Source was supported by the Office of Basic Energy Sciences of the U.S. Department of Energy under Contract No. W-31-109-Eng-38.

\pagebreak

\pagebreak

\noindent {\bf Figure Captions}

\vspace{0.25in}

\noindent Figure 1:  Resistivity as a function of reduced temperature ($\textstyle t\equiv\frac{T-T_{\rho/co}}{T_{\rho/co}}$) in La$_{0.7}$Ca$_{0.3}$MnO$_{3}$ (solid) and Pr$_{0.7}$Ca$_{0.3}$MnO$_{3}$ (dashed).  Note that the Pr$_{0.7}$Ca$_{0.3}$MnO$_{3}$ data were obtained using a 2-point probe measurement of the resistance, and include an estimated scale factor for the conversion to resistivity.

\noindent Figure 2:  Reciprocal space scans along H (a) and K (b), with Lorentzian-squared fits (line), in La$_{0.7}$Ca$_{0.3}$MnO$_{3}$.  Data were measured at temperatures of 260 K (open) and 220 K (closed).  The peaks at (1.67 2 0) and (1.5 2.13 0) arise from powder lines and were excluded from the fits.  Inset of (a) shows temperature dependence of scattering intensity at (1.5 2 0) (open) and (1.3 2 0) (closed).  Note that spurious points at temperatures between 270 and 275 K coincided with a beam dump.

\noindent Figure 3:  Temperature dependence of the scattering at (1.5 2 0), shown with Lorentzian-squared fits (solid), in Pr$_{0.7}$Ca$_{0.3}$MnO$_{3}$.  For clarity, each data set and fit are shifted upward by 125 s$^{-1}$ with respect to the next higher temperature data set and fit.

\noindent Figure 4:  Peak intensity of the (1.5 2 0) scattering--- normalized to the (2 2 0) Bragg peak intensity--- as a function of reduced temperature ($\textstyle t\equiv\frac{T-T_{\rho/co}}{T_{\rho/co}}$), in La$_{0.7}$Ca$_{0.3}$MnO$_{3}$ (open) and Pr$_{0.7}$Ca$_{0.3}$MnO$_{3}$ (closed).

\noindent Figure 5:  The fitted HWHM values of the (1.5 2 0) scattering as a function of reduced temperature ($\textstyle t\equiv\frac{T-T_{\rho/co}}{T_{\rho/co}}$), in La$_{0.7}$Ca$_{0.3}$MnO$_{3}$ (open) and Pr$_{0.7}$Ca$_{0.3}$MnO$_{3}$ (closed).

\noindent Figure 6:  Schematic diagram of the structure of a CE-type bipolaron, in the $a-b$ plane.  Open circles represent Mn$^{4+}$ ions; elongated figure-eights represent the occupied e$_{g}$ (3d$_{z^{2}-r^{2}}$) orbital of Mn$^{3+}$ ions; closed circles represent Mn ions that, on average, have the formal valence and no net orbital order; and arrows indicate the in-plane component of the magnetic moment.

\begin{figure}
\epsfxsize=.5\textwidth
\centerline{\epsffile{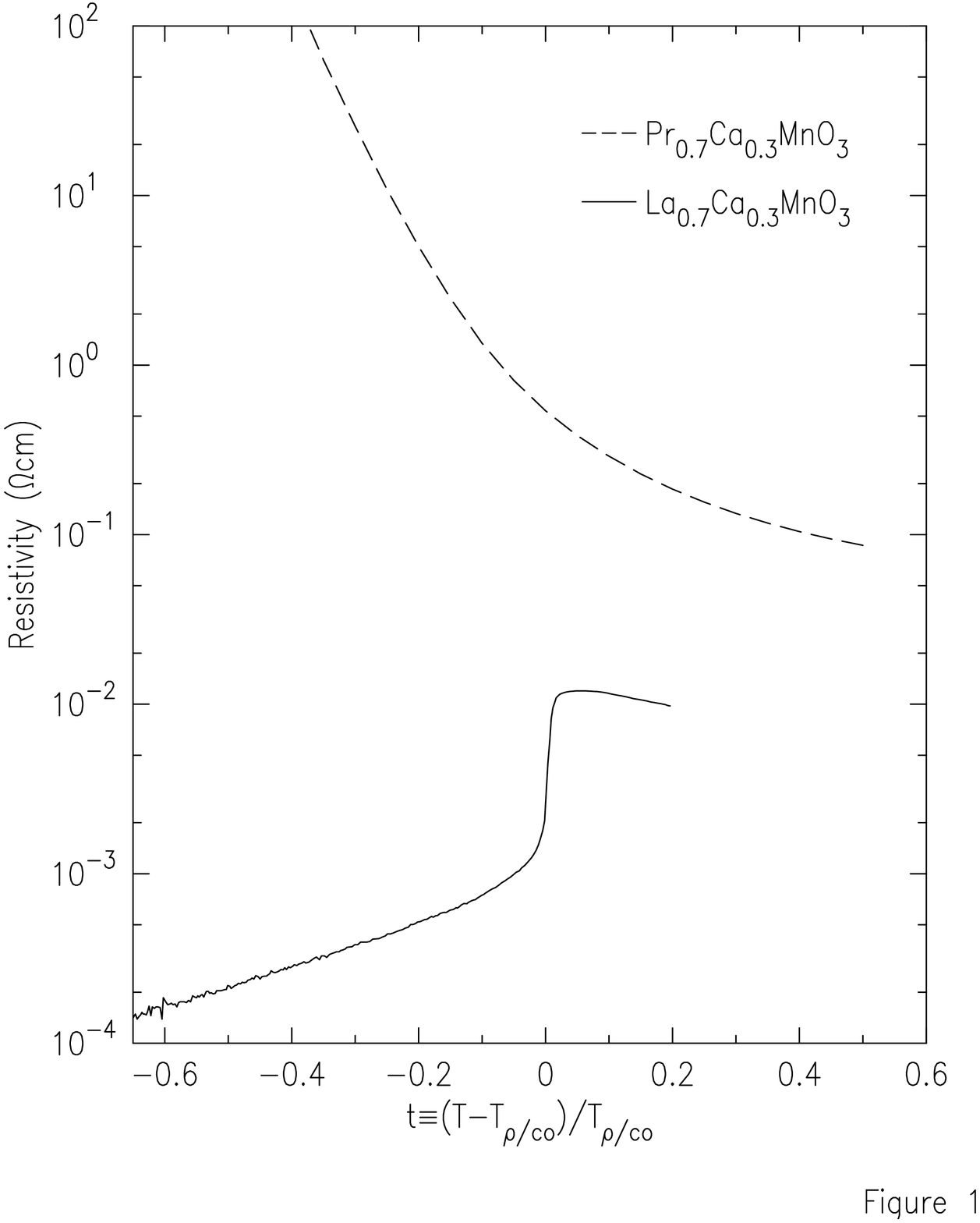}}
\end{figure}

\begin{figure}
\epsfxsize=.5\textwidth
\centerline{\epsffile{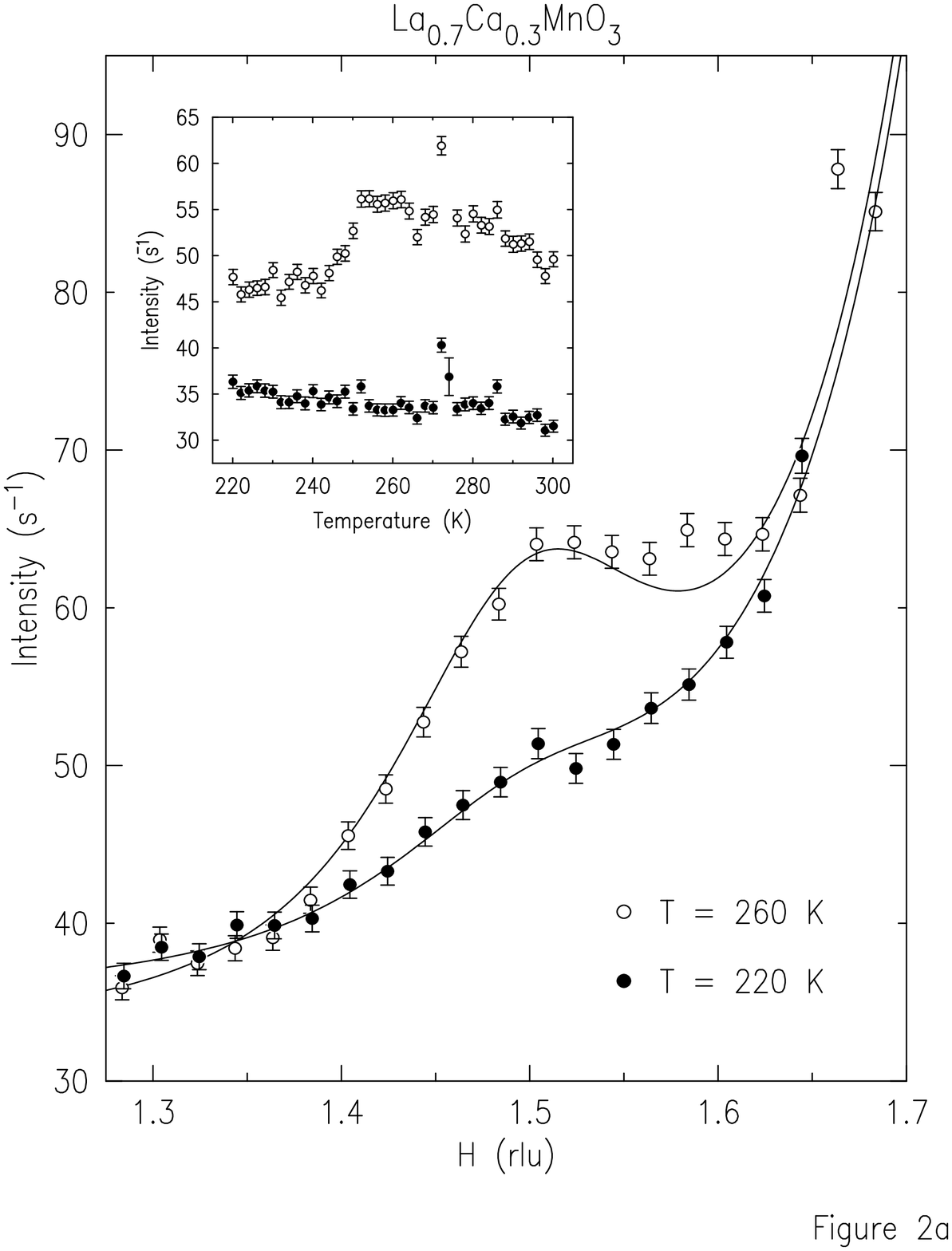}}
\end{figure}

\begin{figure}
\epsfxsize=.5\textwidth
\centerline{\epsffile{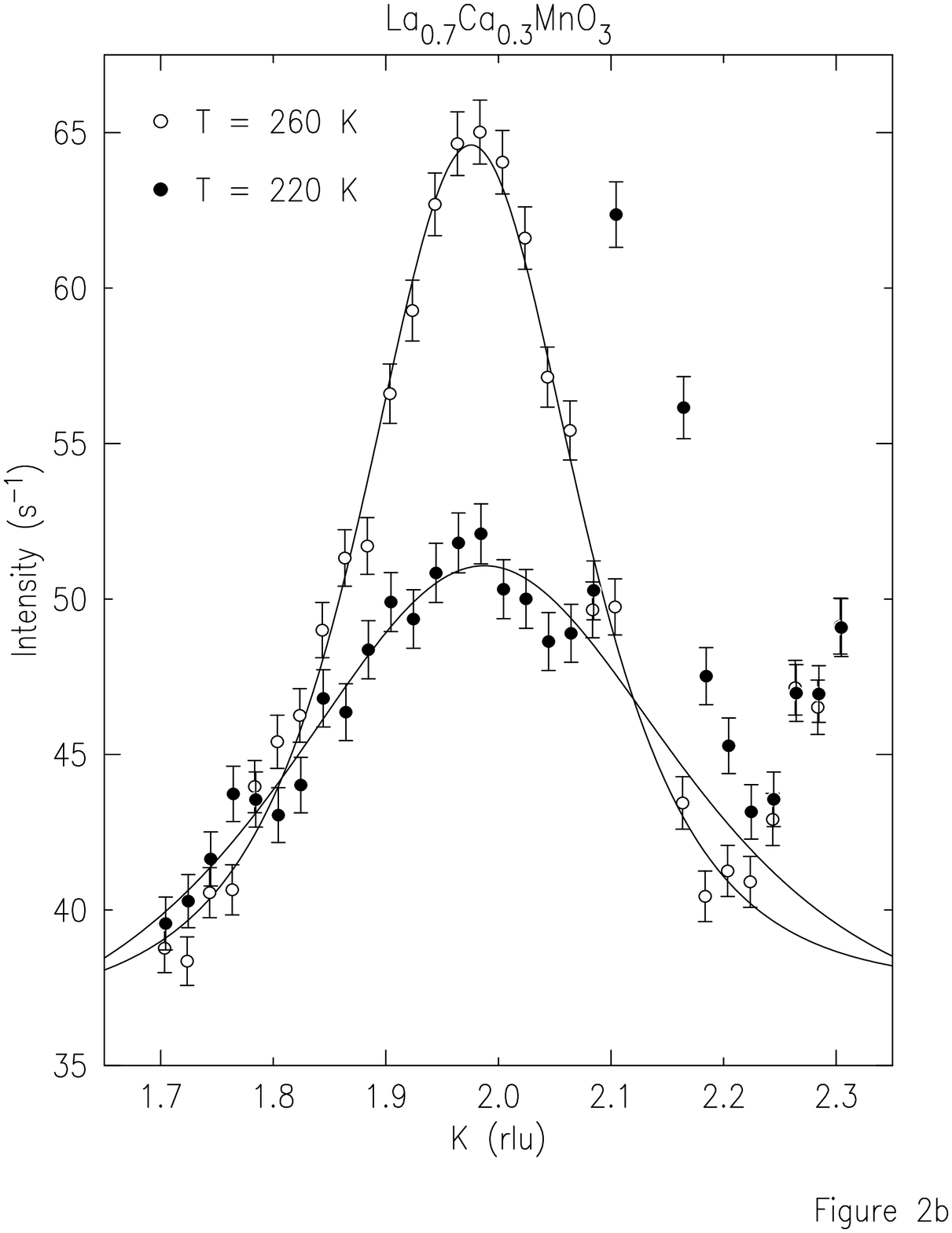}}
\end{figure}

\begin{figure}
\epsfxsize=.5\textwidth
\centerline{\epsffile{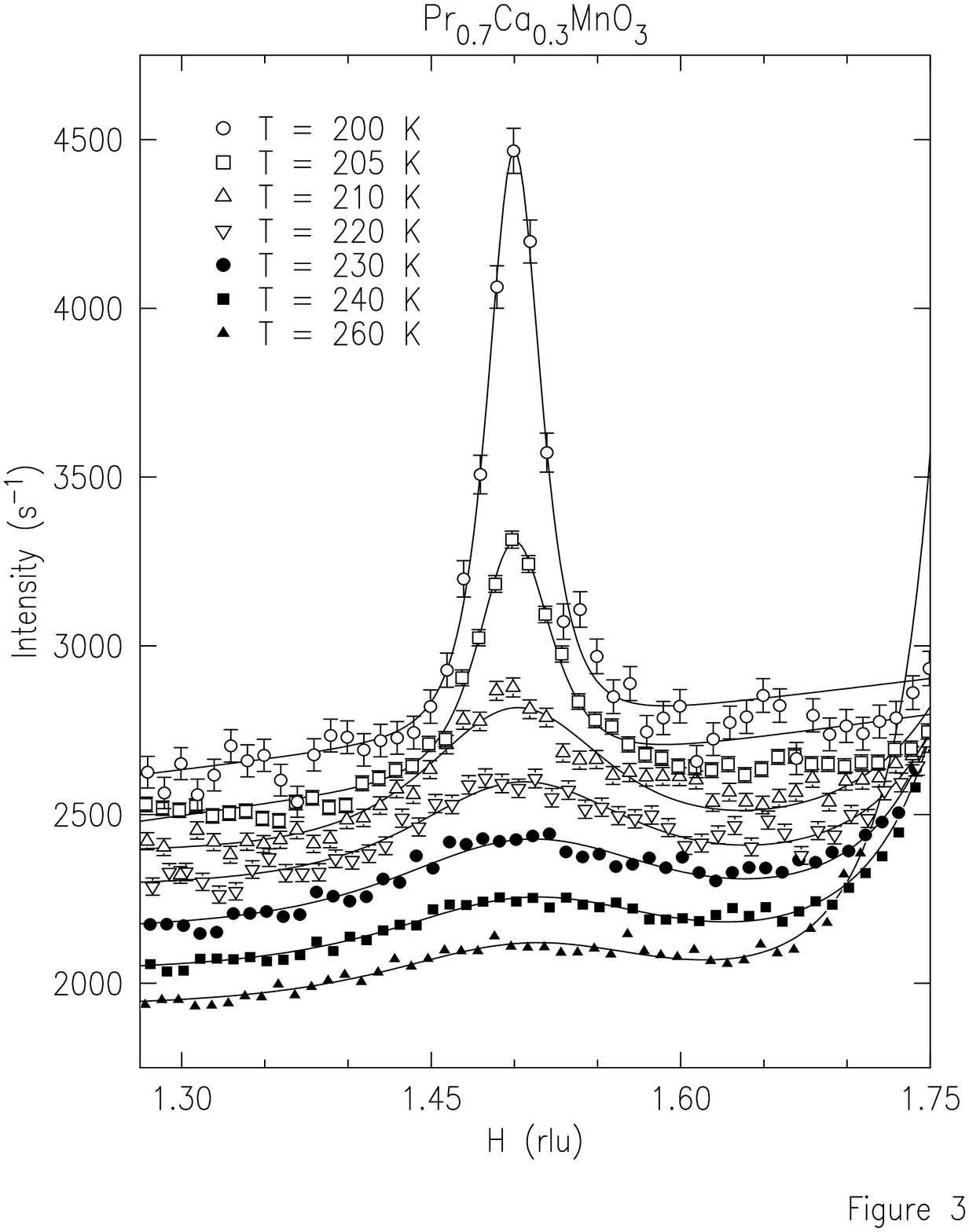}}
\end{figure}

\begin{figure}
\epsfxsize=.5\textwidth
\centerline{\epsffile{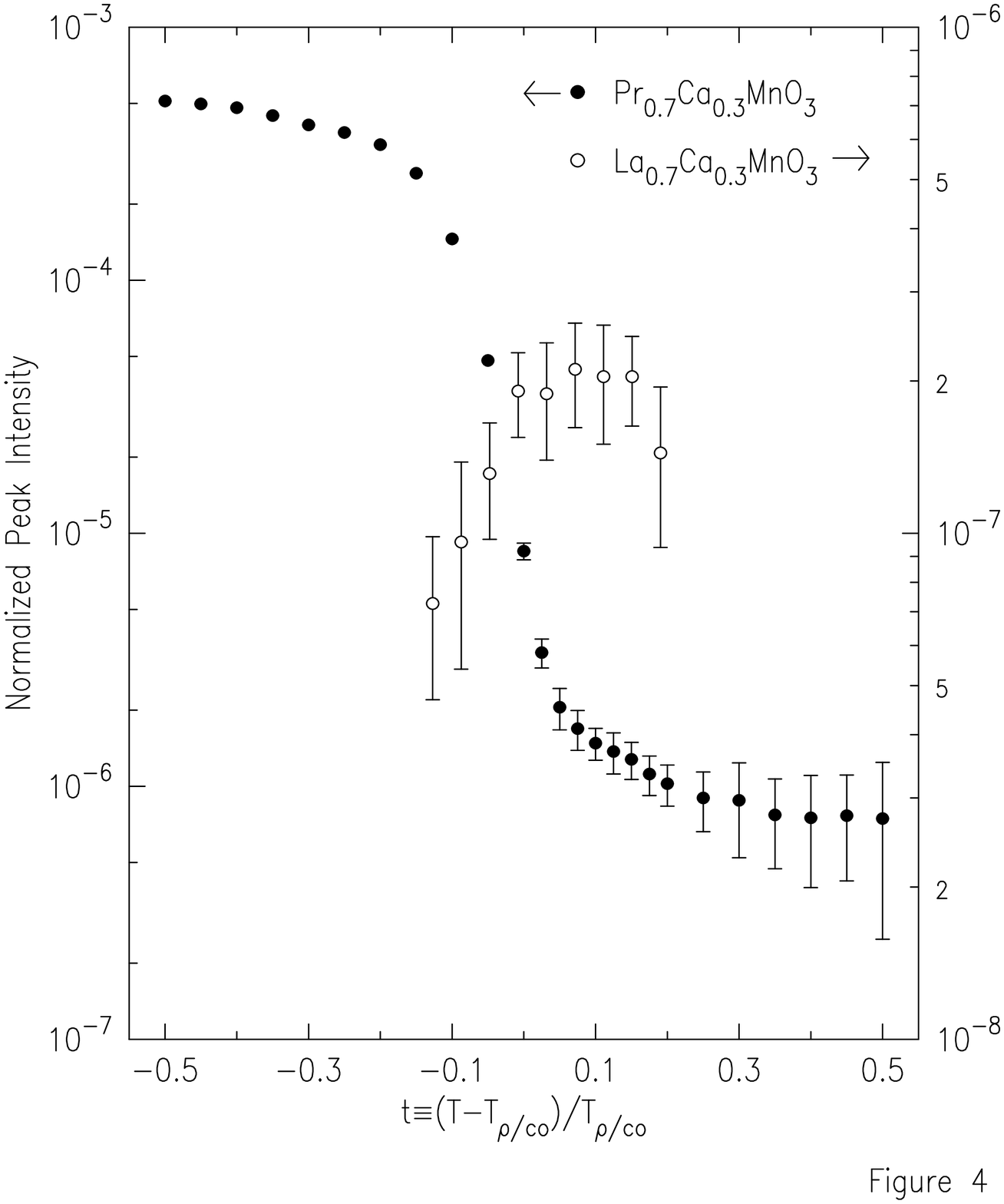}}
\end{figure}

\begin{figure}
\epsfxsize=.5\textwidth
\centerline{\epsffile{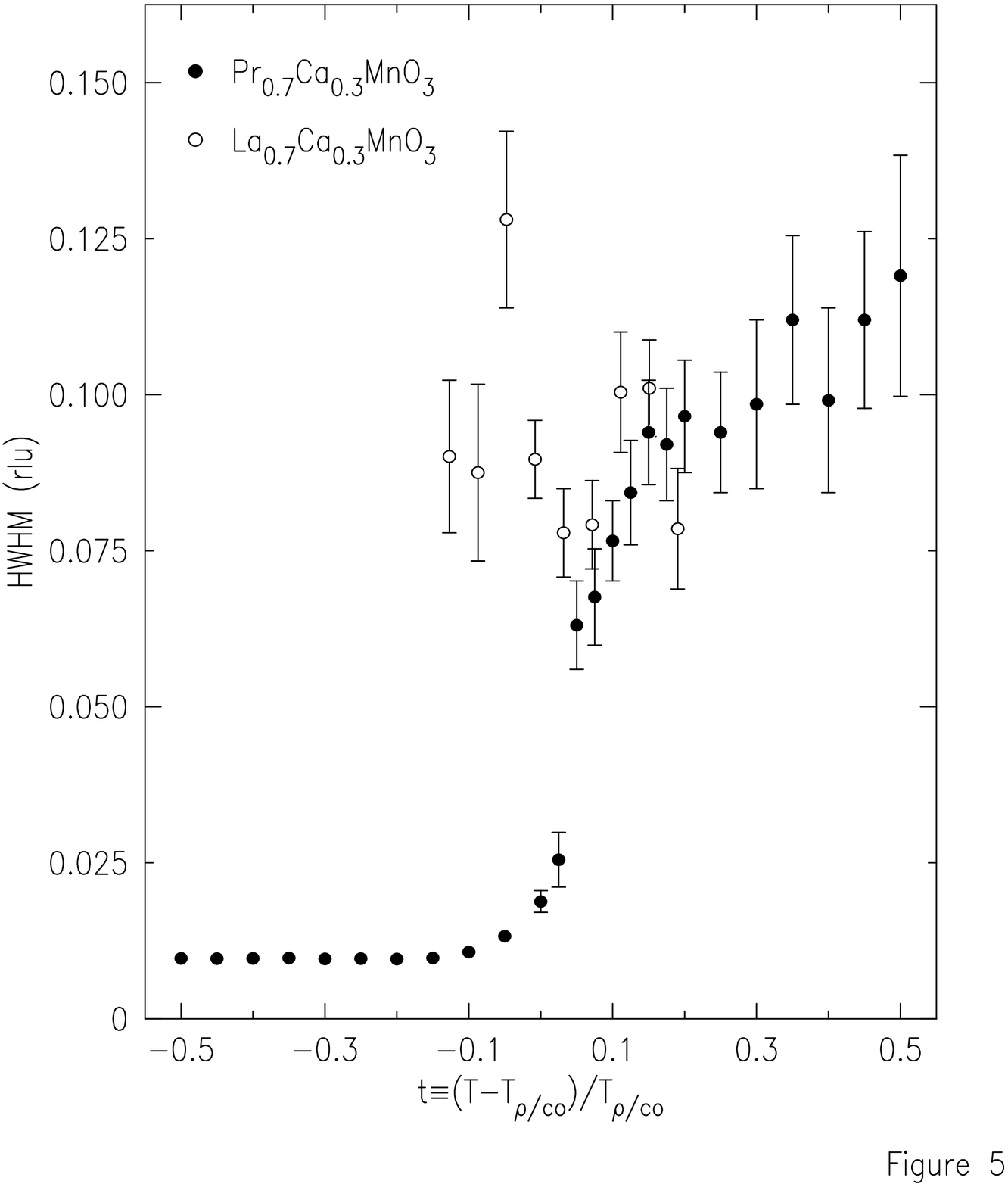}}
\end{figure}

\begin{figure}
\epsfxsize=.75\textwidth
\centerline{\epsffile{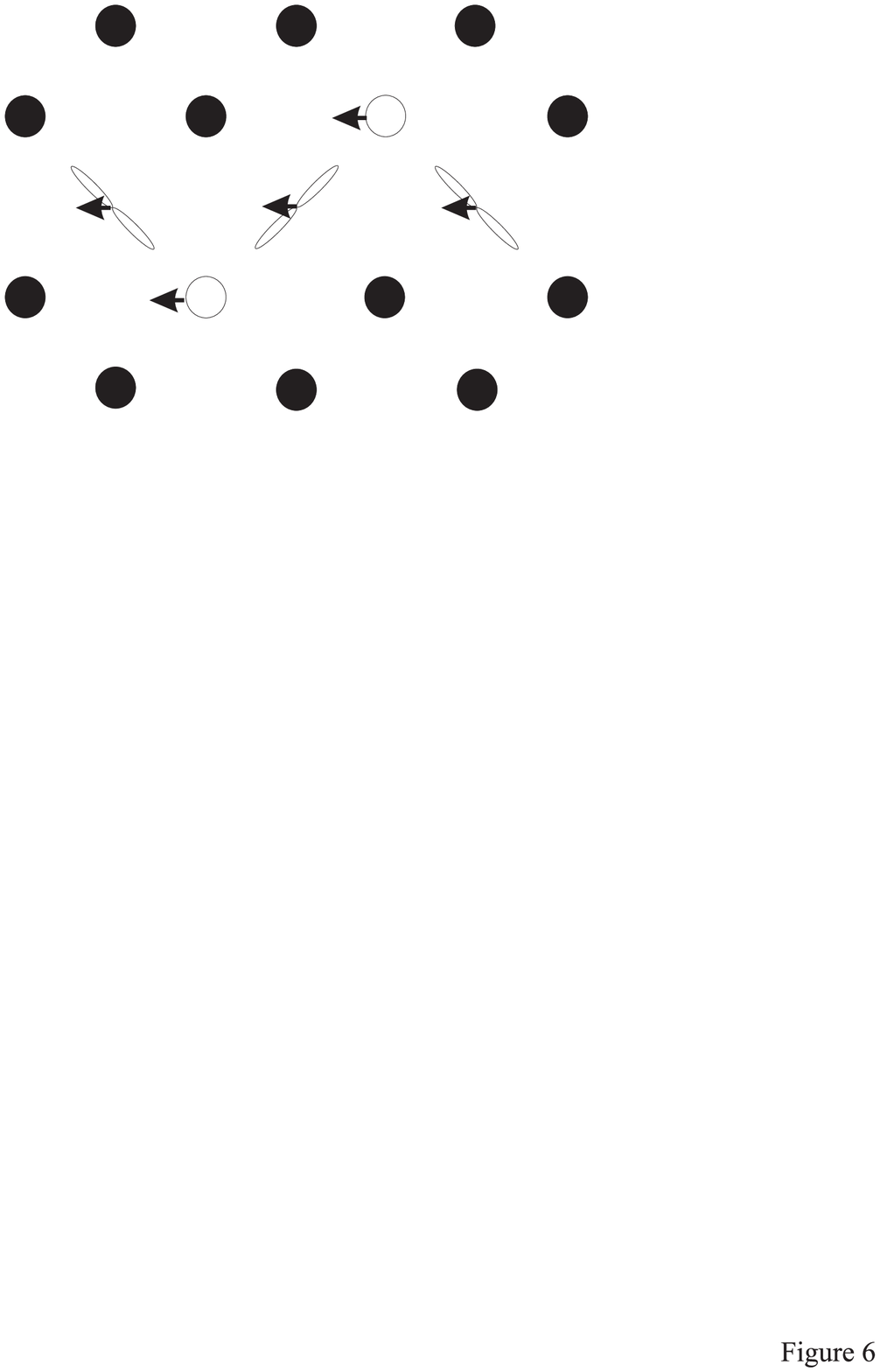}}
\end{figure}

\begin{thebibliography}{99}
\bibitem{caveat} Note that the reciprocal space mesh scans were carried out around the (220) Bragg peak using an incident x-ray energy near the Mn K edge (6.539 keV) and a graphite analyzer.  Previous reports \cite{shimo,doloc,shimo2} of x-ray diffuse scattering attributed to the presence of single polarons utilized high-energy x-rays, which enable scattering at larger Q.
\bibitem{nojust} The Lorentzian-squared lineshape was chosen for the quality of the fit, and no theoretical justification is provided.
\end{thebibliography}
\end{document}